# Application of magnetically induced hyperthermia on the model protozoan Crithidia fasciculata as a potential therapy against parasitic infections

V. Grazú[1]
A.M. Silber[2]
M. Moros[1]
L. Asín[1]
T.E. Torres[1,3,4]
C. Marquina[3,4]
M.R. Ibarra[1,3]
G.F. Goya[1,3]

[1]Instituto de Nanociencia de Aragón (INA),Universidad de Zaragoza, Zaragoza, Spain.
[2]Departamento de Parasitologia, Instituto de Ciências Biomédicas, Universidade de São Paulo, São Paulo, Brazil.
[3]Departamento de Física de la Materia Condensada, Facultad de Ciencias, Universidad de Zaragoza, Spain.
[4]Instituto de Ciencia de Materiales de Aragón (ICMA), CSIC, Universidad de Zaragoza, Zaragoza, Spain.

Correspondence: Gerardo F. Goya.
Instituto de Nanociencia de Aragón (INA).
Universidad de Zaragoza
Ed. I+D - Calle Mariano Esquillor s/n
Campus Rio Ebro
50018-Zaragoza - Spain
Tel: (+34) 876 555 362
Fax: (+34) 976 76 2776

Email : goya@unizar.es

**Abstract:**

**Purpose:** Magnetic hyperthermia is currently an EU-approved clinical therapy against tumor cells that uses magnetic nanoparticles under a time varying magnetic field (TVMF). The same basic principle seems promising against trypanosomatids causing Chagas´ disease and sleeping sickness, since therapeutic drugs available display severe side effects and drug-resistant strains. However, no applications of this strategy against protozoan-induced diseases have been reported so far. In the present study, Crithidia fasciculata, a widely used model for therapeutic strategies against pathogenic trypanosomatids, was targeted with $Fe_3O_4$ magnetic nanoparticles (MNPs) in order to remotely provoke cell death using TVMFs.

**Methods:** Iron-oxide MNPs with average sizes of <d> ≈ 30 nm were synthesized using a precipitation of $FeSO_4$ in basic medium. The MNPs were added to **C**rithidia fasciculata choanomastigotes in exponential phase and incubated overnight, removing the excess of MNPs by a DEAE-cellulose resin column. The amount of uploaded MNPs *per cell* was determined by magnetic measurements. The MNP-bearing cells were submitted to TVMFs using a homemade AC-field applicator ($f = 249$ kHz, H = 13 kA/m), and the temperature variation during the experiments was measured. Scanning Electron Microscopy was used to asses morphological changes after TVMF experiments. Cell viability was analyzed using the MTT colorimetric assay and flow cytometry.

**Results:** The MNPs were incorporated by the cells, with no noticeable cell-toxicity effects. When a TVMF was applied to MNP-bearing cells, massive cell death was induced via a non-apoptotic mechanism. No effects were observed by applying a TVMF on control (without loaded MNPs) cells. No macroscopic rise in temperature was observed in the extracellular medium during the experiments.

**Conclusion:** These data indicate (as a proof of principle) that intracellular hyperthermia is a suitable technology to induce the specific death of protozoan parasites bearing MNPs. These findings expand the possibilities for new therapeutic strategies that combat parasitic infections.

# Introduction

Diseases caused by organisms of the *Trypanosoma* and *Leishmania* genera affect a global population of at least 20 million people, with an estimated at risk population of approximately 450 million people [1,2]. These statistics indicate that trypanosomatid-induced diseases are a severe sanitary problem, with the resulting disease burden affecting much of the population that resides in the tropical and subtropical regions of the globe. [3,4] Despite the sanitary relevance of trypanosomatid-induced diseases for human health, no satisfactory treatments exist to combat these infections. [5-7] The two therapeutic agents that are presently in use for the treatment of Chagas´ disease are Nifutrimox and Benznidazole; however, these drugs were developed approximately 40 years ago [8]. The main disadvantages of these treatments are a high level of toxicity and a low therapeutic efficiency during the chronic phase of the disease. The latter disadvantage is a serious problem because Chagas´ disease is mainly diagnosed during the chronic phase; therefore, the majority of infected people miss the opportunity to be treated using effective chemotherapy [9]. In addition, several cases of drug-resistant or partially resistant strains have been reported for both of these drugs [10]. The initial stages of *Trypanosoma brucei* infections, when the central nervous system is not compromised, can be treated using suramin or pentamidine [11]. Again, these drugs are not effective during the late stages of the disease (when the majority of cases are diagnosed) because these drugs do not traverse the brain blood barrier. For these cases the first-line treatment is melarsoprol, a drug that can cross the blood-brain barrier. Melarsoprol is a highly toxic drug that causes a myriad of serious undesired side effects, including reactive encephalopathy, in approximately 20% of the patients receiving treatment [12]. For the treatment of diseases caused by *Leishmania* spp., Pentavalent antimonials are most often used [13]. Although these drugs are effective for treating the cutaneous form of the disease, treatments for the parenteral form of the disease are limited. In addition, these drugs are highly toxic. Amphotericin B and pentamidine are considered to be second-line drugs because of their serious or irreversible toxic effects. However, these drugs are now being reconsidered on the basis of new formulations or dosage regimens [13-15]. The fact that the majority of the drugs that are currently used for trypanosomatid-induced diseases were developed approximately 40 years ago reveals the limited success of the strategies to develop novel therapeutic treatments. This lack of success highlights the necessity for new strategies and tools to address this important public health issue.

Magnetic hyperthermia is a relatively new medical protocol [16] that uses magnetic nanoparticles to heat areas of the body using the application of time-varying magnetic fields (TVMFs). The physical mechanisms underlying energy absorption by MNPs are related to the existence of the magnetic relaxation of single domains by Arrhenius-Néel processes [4,17]. With the advent of nanotechnology, it became possible to engineer efficient MNPs that have the ability to absorb large amounts of energy from TVMFs (up to several kW per gram of material) to induce a local rise in temperature. [18] Because of their size, these nanoparticles can be incorporated inside target cells, making possible to heat up small foci at the single cell level.[19] The application of hyperthermia using MNPs, alone or in combination with other therapeutic strategies, was proposed more than a decade ago as a therapeutic technique to treat cancer.[16] In the present study, we used the non-pathogenic trypanosomatid *Crithidia fasciculata*, a well-accepted model of other pathogenic trypanosomatid parasites [20], to evaluate the use of magnetic hyperthermia as a potential trypanocidal treatment. The mechanisms of cell death and the application of these principles of magnetic hyperthermia to treat parasitic diseases are discussed in later sections of the present work.

# Material and methods

## *Reagents and Culture media*

The chemicals and fetal calf serum (FCS) used for the present study were purchased from Sigma (Missouri – USA). The other components of the culture medium were purchased from Difco (Lawrence-USA). The apoptosis detection kit was purchased from Immunostep (Coimbra – Portugal). The diethylaminoethyl-cellulose (DEAE-cellulose, DE52) was purchased from Whatman (Dassel- Germany).

## *Cells*

*Crithidia fasciculata* choanomastigotes were grown at 28ºC in a Warren culture medium [37 g/l brain heart infusion, 100 ng/l hemin, 100 mg/l folic acid] supplemented with 10% fetal calf serum. The cells were seeded in 75 cm$^3$ tissue culture plates at 1 x 10$^6$ cells/ml. *C. fasciculate* cultures were incubated until they reached the exponential phase (approximately 24 h of incubation). Using daily subculturing, the cells were maintained in the exponential growth phase for use in the studies. Cell counting was performed in a Neubauer chamber. The cells were evaluated for viability using the optical microscopic observation of flagellar motility and by counting of the number of viable cells after incubation with 2% trypan blue in PBS.

## Magnetic Nanoparticles

The MNPs used in the present study were synthesized using a precipitation of iron (II) salt ($FeSO_4$) in the presence of a base (NaOH) and a mild oxidant ($KNO_3$) under a Nitrogen atmosphere, as previously described in the literature [21]. Mixing the reactants during a 24 h period resulted in $Fe_3O_4$ particles with average sizes of <d> ≈ 30 nm and with a colloidal stability in aqueous medium at pH 7.

## Cell uptake of MNPs and separation of the non-incorporated nanoparticles.

The cells collected during exponential growth phase were centrifuged and resuspended in fresh culture medium, adjusting concentrations to $2.5 \times 10^8$ cells/ml. The MNPs were added to a final concentration of 0.425 mg/ml and were incubated at $28^oC$ overnight with gentle agitation. To separate the cells from the non-incorporated MNPs, we took advantage of the fact that MNPs (isoelectric point= 5.0) adsorb to DEAE-cellulose resin at pH 7.0 (isoelectric point = 2.5), whereas *C. fasciculata* does not interact with the resin under these conditions. Briefly, the cells were washed twice with PBS, resuspended in 12 ml of 2% glucose in PBS (gPBS), and incubated with 6 g DEAE-cellulose ionic exchange resin that was previously equilibrated with gPBS for 10 min with gentle agitation at room temperature. The cells were recovered with an efficiency of more than 95%.

## Determination of cell-incorporated MNPs by magnetization measurements

The amounts of cell-associated MNPs were determined by measuring the saturation magnetization using a Superconducting Quantum Interference Device (SQUID) (MPMS-7T, Quantum

Design). The magnetization measurements were conducted on dried MNPs or lyophilized MNP-loaded cells and were performed as a function of the applied magnetic field up to 5 kOe (0.4 MA/m) at different temperatures between 5 and 300 K.

## Experiments of time-varying magnetic field application

Aliquots of MNPs (10 mg/ml) or MNP-bearing cells ($1.7 \times 10^9$ cells/ml) were submitted to alternating magnetic fields using a homemade AC-field applicator. The magnetic field applicator, consisting of a resonant LC tank working close to the resonant frequency, was used to measure the specific power absorption (SPA) of the samples. A magnetic field ($f = 249$ kHz, H = 13 kA/m) was achieved inside a gap of four (2+2) turns of a copper tube around high-permeability polar pieces. The SPA values were obtained from adiabatic measurements inside an insulated Dewar. The temperature data was measured using a fiber-optic temperature probe (Reflex[TM], Neoptix) that was immune to the radiofrequency environment. Prior to each experiment, the temperature evolution was measured between the 5 and 10 min time points with the RF-source turned off to establish a T-baseline. Next, the power was turned on, and the temperature rise was monitored for a 30 min interval.

## Scanning Electron Microscopy (SEM)

Cells in the exponential growth phase were fixed with 2.5 % glutaraldheyde in 0.1 M sodium cacodylate 3% sucrose solution for 90 min at 4ºC. The dehydration process was conducted by incubating the cells in increasing concentrations of methanol at 30, 50, 70, and 100%. Each of these concentrations were used for 5 minutes in duplicate, and a final step was conducted using anhydride methanol for 10 minutes. A drop of the dehydrated cells in suspension was placed over a coverslip. Next, when the methanol was evaporated, the coverslip was gold coated. The samples were then observed using scanning electron microscopy (EDX Hitachi S-3400 N, Instituto Carboquimica CSIC). Secondary electron images were also performed.

## *Viability analysis*

## MTT assay

Cell viability was analyzed using the MTT colorimetric assay. For the cytotoxicity assay, $5x10^6$ cells (MNPs$^-$/TVMF$^-$, MNPs$^-$/TVMF$^+$, MNPs$^+$/TVMF$^-$ or MNPs$^+$/TVMF$^+$) were resuspended in 100 µl of Warren culture medium. Next, 40 µL of MTT dye solution (5 mg/mL in PBS) was added to each aliquot. After 4 h of incubation in eppendorf tubes at 28°C, formazan crystals were dissolved by the addition of 100 µL of 10% SDS. All of the cell debris, which has been shown to interfere with the assay, was removed by centrifugation (10 min at 12000 r.p.m.). Next, the absorbance of each supernatant was read using a microplate reader (Biotek ELX800) at 570 nm. The spectrophotometer was calibrated to zero absorbance using a culture medium without cells. The relative cell viability (%) compared with the control cells (the exponential-phase cells not submitted to any treatment) was calculated by [Absorbance]test/[Absorbance]control x 100. Each measurement was repeated at least five times to obtain mean values with standard deviations.

## Flow cytometry

The cell viability was also measured using flow cytometry with the commercial apoptosis detection kit purchased from Immunostep. Briefly, $1x10^6$ cells of each sample were resuspended in Annexin-binding buffer, were stained with 5 ㎖ of Annexin and 5 ㎖ of propidium iodide, and were incubated for 15 minutes at room temperature in the dark. The cell analysis was performed using the FACSAria Cytometer (Becktom Dickinson) and FACSDiva Software.

# Results and discussion

The magnetic and physicochemical properties of the MNPs used in the present study were reported elsewhere.[22] The magnetic colloids were composed of cubic $Fe_3O_4$ nanoparticles of average size <d> = 30 ± 8 nm with saturation magnetization at room temperature of 85 emu/g, which is close to that of the bulk magnetite magnetization.[23] This synthesis route resulted in magnetic colloids with isoelectric point of 5.0, electrostatically stabilized due to the adsorption of $SO_4^=$ groups on the particle surface. The ability to dissipate heat under a TVMF of any type of MNPs is measured by the specific power absorption (SPA) given in watts per gram of magnetic material, which for the present MNPs and experimental conditions ($f$ = 249 kHz, H = 13 kA/m) was found to be 83.6 $W/g_s$[22] comparable to other SPA values reported in the literature for $Fe_3O_4$ particles of similar size.[24-26] The selection of MNPs was made based on the known dependence of the SPA values on the average particle size and size distribution. Indeed, the strong dependence of the Nèel relaxation-based model on particle size yields a maximum value of SPA for magnetite MNPs within a narrow range of diameters <d> around 15-30 nm,[4] as experimentally confirmed in many colloidal systems yields. [3,19,21] The precise value for this maximum will depend on other magnetic properties of the MNPs such as magnetic anisotropy and saturation magnetization.

The optimal experimental conditions for the *C. fasciculata* to incorporate the MNPs were determined by a series of experiments as a function of the incubation time, incubating *C. fasciculata* with a fixed MNP concentration for increasing times from 15 to 240 min. After the incubation, the non-incorporated MNPs were separated by the column method already detailed in the Materials and Methods section. The incorporated mass of magnetic material per cell was determined by measuring the saturation magnetization $M_S$ of 100 ㎖ of culture medium containing ≈$10^9$ cells and comparing these values with the $M_S$ of the pure colloid.[27] From these data and the average particle size, the average number of incorporated MNPs per cell was calculated (Table 1). The number of incorporated MNPs decreased from a maximum of approximately $10^8$ NPs/cell after 15 min of incubation to approximately $10^6$ NPs/cell after 1 h of incubation. Shorter incubation times resulted in a low and highly variable cell charging, possibly due to the fact that the cells require an induction time to activate the biological mechanisms to incorporate the MNPs. The time course of this decrease of the cell-associated MNPs followed an exponential decay ($R^2$=0.9998), reaching a near-steady state after approximately 1 h incubation, which remained essentially constant up to 12 h of incubation (Fig. 1 and Fig. S1 in the supplementary material). From these experiments, we established the 15 min incubation as being the optimal condition for charging the cells with a reproducible and defined number of MNPs/cell.

To evaluate the suitability of the MNPs for hyperthermia applications, it was necessary to assess the influence of these NPs on cell viability and on the incorporation and separation conditions. It was observed that, when compared with exponential-phase cells, the NP-treated or mock-treated cultures that were submitted to separation conditions (incubated with DEAE-cellulose resin in gPBS) remained more than 95% viable (data not shown). This finding illustrated that the observed cell death in the subsequent experiments was not due to the previous exposure of the cells to toxic conditions.

## *Hyperthermia experiments*

After the best conditions for maximum uptake of MNPs by the cells were determined (15 min incubation, as stated above), the hyperthermia experiments were performed. The experiments were designed in a 2x2-row-column format (Fig. 2), in which the four groups were defined as follows: **a.** cells not-bearing MNPs that were not-submitted to TVMFs ($MNPs^-/TVMF^-$); **b.** cells bearing MNPs that were not-submitted to the magnetic field application ($MNPs^+/TVMF^-$); **c.** cells not-bearing the MNPs that were submitted to the magnetic field application ($MNPs^-/TVMF^+$); and **d.** cells bearing the MNPs that were submitted to the magnetic field application ($MNPs^+/TVMF^+$). The effects on cell viability of the above mentioned treatments were evaluated using several criteria, including the direct observation of the cell motility using optical microscopic observation, the mitochondrial activity using a MTT assay, and the detailed morphology using scanning electronic microscopy (SEM). The qualitative observation of the samples that were examined using the previously described treatments revealed that the $MNPs^+/TVMF^+$ population caused 100% cell death (Fig. 2, supplementary material). The quantitative analysis of the cells using a MTT assay revealed that the cell viability of all of the other samples was not affected compared with the control ($MNPs^-/TVMF^-$) (Fig. 3). The analysis of all four samples using SEM revealed severe structural damage in the cell morphology only for the $MNPs^+/TVMF^+$ population, particularly at the level of the cell surface, indicating severe plasma membrane damage (Fig. 4).

## *Cell death*

Because the application of the TVMFs was performed in an adiabatic device, the hyperthermia treatment could have produced a transient macroscopic increase of the sample temperature due to differences in the rates of heat generation and dissipation. When a positive control experiment was conducted by applying a TVMF to a suspension that only contained MNPs (Fig. 5), the sample showed a large increase of temperature, of about 50ºC along the 30 minutes of the experiment. Therefore, we tested whether a similar increase in average temperature might have contributed to the amount of cell death by monitoring the temperature of the extracellular medium during the application of the TVMF. The results on both control and magnetically loaded cells showed only a slight macroscopic increase in temperature (about 2-4 ºC) after 30 min of TVMF application. This result (i.e., the absence of temperature increase in the samples composed of magnetically-loaded cells) is the expected based on the much lower 'average concentration' of MNPs in these samples, since the small amounts of uploaded MNPs are contained within a total volume of about 0.5 ml of liquid cell medium.

These results clearly demonstrate that the heat released from the MNPs was not enough to increase the average temperature of the cell culture in such a way that would compromise the viability of the cells. Therefore, the origin of the cell death that was measured after the application of the TVMF should not be related to thermal stress. This is in agreement with previous works on magnetically-loaded human dendritic cells[19,28], which demonstrated that application of TVMF for 30 min yielded up to 90-95 % of cell death, without affecting blank (i.e., without MNPs) cells. Similar reports have been reported in HeLa cell line [29] loaded with MNPs. Some theoretical models on metal nanoparticles have suggested this possibility.[30] Since the temperature was essentially constant during experiments, the observed cell death suggests an intracellular, MNP-triggered mechanism different from the temperature-induced apoptosis by hyperthermia. However, as the temperature was measured with a macroscopic sensor, the possibility of intracellular heating up to apoptotic temperatures cannot be excluded.

To evaluate the plausibility of intracellular heating during our experiments, a simplified heat transfer model at the single-cell scale was considered, e.g. a single cell magnetically loaded with MNPs, surrounded by a large matrix of unloaded cells (or just culture medium free of MNPs). We estimated the expected intracellular temperature increase in such a case based on a) the measured mass of MNPs at the intracellular medium; b) the calculated average cell volume from SEM images ($230\pm50$ $mm^3$), and c) the measured SPA values of the pure magnetic colloid. For simplicity, we further approximated the specific heat capacity of a single cell to the pure water value $C_P = 4.18$ J/(g.K). From the average cell volume and MNPs upload (1-10 pg/cell), we estimated a temperature increase rate of 0.02-0.87 K/s. The above calculations were made considering that no heat was dissipated from the intracellular medium to the cell environment, a clearly unrealistic hypothesis. Since the cell membrane has a non-negligible thermal conductivity, the calculated heating rates are not enough to rise the intracellular temperature up to the 41-45 °C needed for triggering thermally induced apoptotic mechanisms.

Several prior studies suggest that apoptosis-related mechanisms are a main cause of hyperthermia-associated cell death [31,19]. Apoptosis (cellular programmed death) is a precise mechanism in which cells follow a programmed sequence of events to induce their death with minimal disturbance to the total cell population. [32] This phenomenon appears to be present in a wide range of organisms, from primitive single-cell to higher multi-cellular eukaryotes. In the present work, we investigated whether cell death by magnetic hyperthermia of *C. fasciculata* was attributable to apoptosis. An early event that is considered as a marker of apoptosis is the exposition of phosphatidylserine on the external surface of the plasma membrane. The cells that incorporated nanoparticles, and the controls, that were either submitted or not submitted to the magnetic field, were incubated with annexin (used to detect the presence of phosphatidylserine) and propidium iodide (used to detect damage to the plasma membrane). Next, these four populations of cells were analyzed using flow cytometry. As illustrated in Fig. 6, the results confirm the 100% viability for the MNPs$^-$/TVMF$^-$, MNPs$^+$/TVMF$^-$ and MNPs$^-$/TVMF$^+$ cell samples and the 0% viability of the MNPs$^+$/TVMF$^+$ cells. In this last case, it was observed that the cells were reactive to annexin and permeable to propidium iodide, indicating plasma membrane damage, which was confirmed using SEM (Fig. 4). These results suggest that the application of a TVMF to cells that have incorporated magnetic nanoparticles results in cell death via a non-apoptotic mechanism. Recent works on the application of TVMF on magnetically-loaded cells showed that a large decrease in cell viability can be achieved without actual temperature increase of the cell medium.[29,33] Furthermore, it has been reported that in the case of magnetically-loaded dendritic cells the percentage of cell death was proportional to the amount of uptaken MNPs. As the cell death observed in our work correspond to the maximum amount of uploaded MNPs (i.e., after 15 min of co-cultivation), it is still to be determined whether a similar effect could be achieved with smaller amounts of uploaded MNPs.

Taken together, our results lead us to propose that in this case, the whole irreversible cell injury due to mechanical stress (evidenced by SEM) in MNPs$^+$/TVMF$^+$ cell samples is the main cause of death. Theoretical calculations on the effect of the power released by MNPs on the cell membrane supports this hypothesis.[34] However, it is worth mentioning that other factors, such as the liberation of toxic proteins into the cytoplasm due to the disruption of membranes compartmentalizing them inside specific organelles like lysosomes, cannot be ruled out as simultaneous cause of cell injury and death.

# Conclusion

In conclusion, the series of experiments reported here demonstrated, as a proof of principle, that magnetically induced hyperthermia causes microorganism death. Our results also illustrate that hyperthermia is a thermal phenomenon of a subcellular scale because no macroscopic increase of temperature was observed. We were able to show that hyperthermia was specific for cells that incorporated MNPs and were submitted to the TVMF because neither NPs nor the TVMF alone resulted in a loss of viability. Lastly, the cells that were submitted to the hyperthermia treatment were dramatically damaged at the plasma membrane level. It should be stressed that although this methodology is being extensively investigated and is currently used for mammalian cells, to our knowledge, this method is not currently being proposed to treat diseases that are caused by microorganisms. The present study highlights this method (at least as a proof of principle) as a potential and novel alternative to treat infections caused by microorganisms. A major advantage of this method is that it causes selective physical damage to the target cells. Therefore, the probability of the emergency of resistance strains is small. More detailed studies are also being conducted to identify the mechanisms involved in cell death at a greater level of detail. Developing delivery systems that have the capability of specifically directing magnetic nanoparticles to infectious agents remains a challenge. These findings lead us to propose this method as a novel strategy to develop new therapeutics against pathogenic microorganisms.

# Acknowledgments/Disclosures

The author reports no conflicts of interest in this work. This work was supported by the Spanish Ministry Ministerio de Ciencia e Innovación (project MAT2010-19326 and Consolider NANOBIOMED CS-27 2006) and IBERCAJA. Partial support from Brazilian grants 08/57596-4 and 11/50631-1 from FAPESP and INBEQMeDI is also acknowledged. We gratefully recognize Dr. M. Vergés, Dr. M.P. Morales and Dr. A.G. Roca for their kind donation of MNPs. We also thank Dr. J. Godino (Instituto Aragonés de Ciencias de la Salud I+CS, Zaragoza) for his help with flow cytometer measurements; I. Echaniz for technical support and L. Casado for helping with scanning electron microscopy imaging.

**Table 1** Average number of MNPs incorporated within a single cell as a function of incubation time, as calculated from the saturation magnetization $M_S$ of the magnetically loaded cells.

| Time (minutes) | 15 min | 30 min | 45 min | 1 h | 1,5 h | 2 h | 4 h | 12 h |
|---|---|---|---|---|---|---|---|---|
| Mass of $Fe_3O_4$ (pg/cell) | 12.3 | 0.55 | 0.31 | 0.14 | 0.22 | 0.15 | 0.23 | 0.18 |
| Number of MNPs ($x10^6$)/cell | 87 | 3.9 | 2.2 | 1.0 | 1.6 | 1.1 | 1.6 | 1.3 |

**Figure 1** Number of MNPs uploaded per cell as a function of incubation time. The observed decrease of incorporated MNPs followed an exponential decay, and reached a near-steady state for incubation times larger than 1 h.

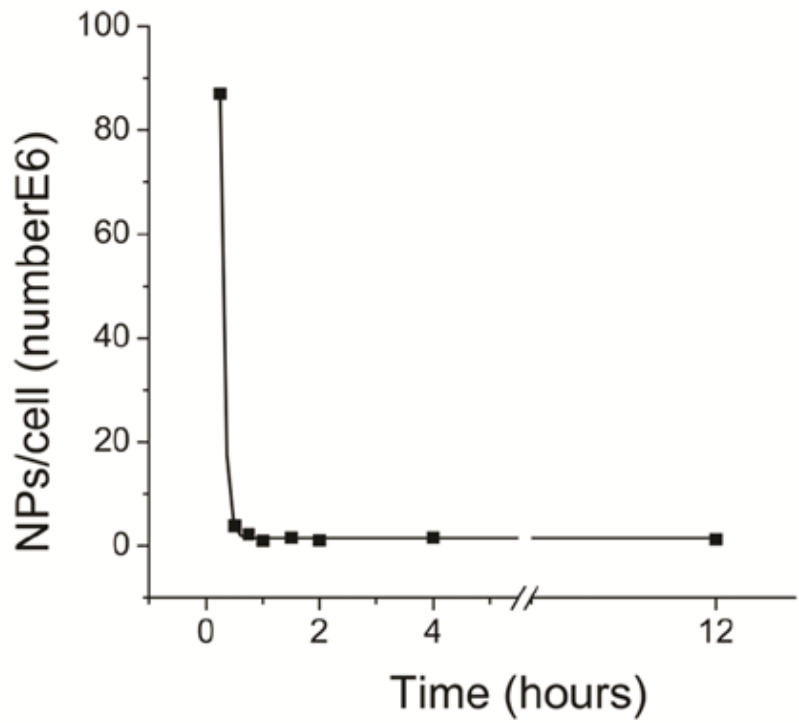

**Figure 2:** Schematic view of the experimental '2x2' design for evaluating the effect of MNPs and TVMF on *C. fasciculata*: a) cells without MNPs not submitted to magnetic fields; b) cells with MNPs without magnetic field application; c) application of magnetic fields on unloaded cells and d) application of magnetic fields on MNP-loaded cells.

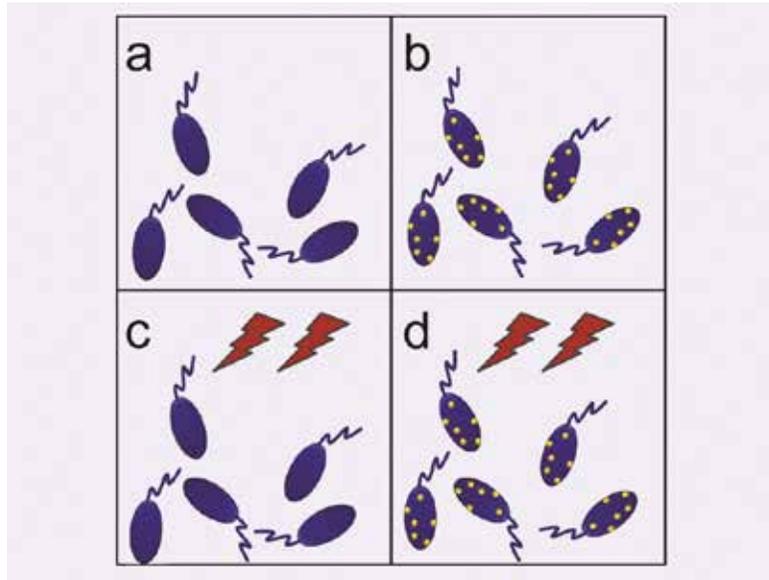

**Figure 3:** MTT results of the four situations displayed in Figure 2. All samples showed 100% of cell viability except in the case of TVMF applied on magnetically loaded cells, where caused 95±5% cell death.

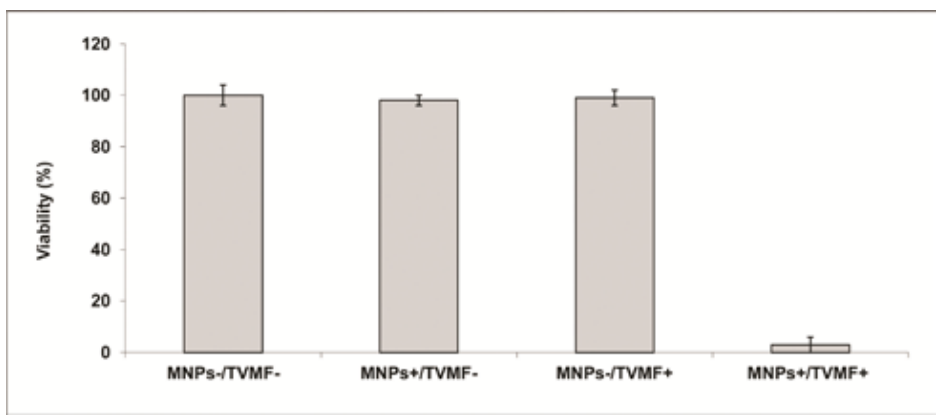

**Figure 4:** SEM images of MNPs⁺/TVMF⁺ sample before (A) and after (B) application of magnetic fields. In the latter case, the changes in cell morphology can be clearly observed, reflecting the severe cell damage after TVMF.

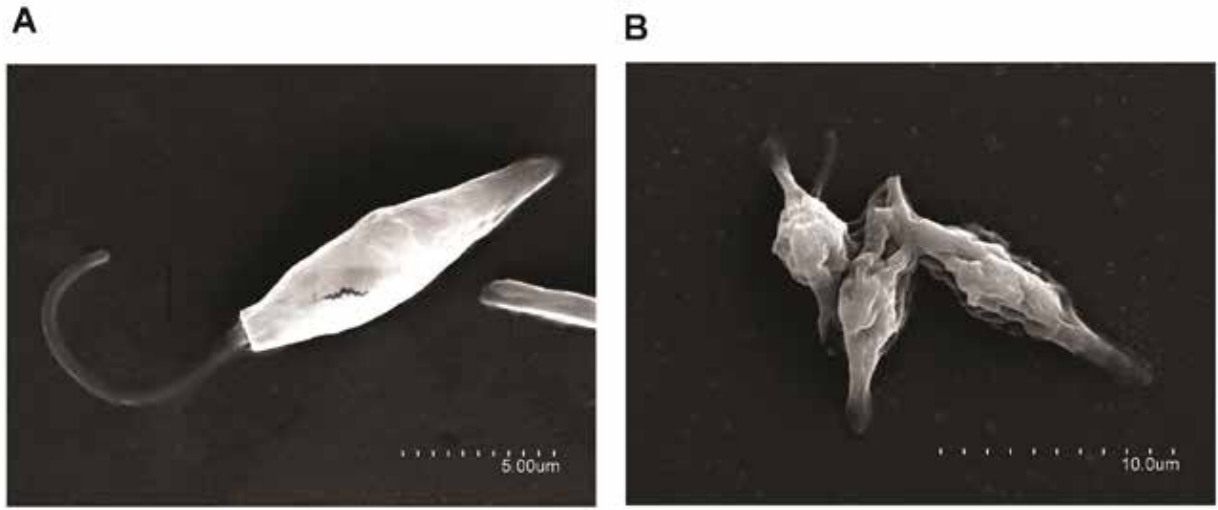

**Figure 5:** Specific Power absorption (SPA) of magnetic colloid (solid squares) at 1 % wt. concentration, and the NP-loaded protozoa (solid line) during application of ac magnetic field (H = 160 Oe, f = 250 kHz).

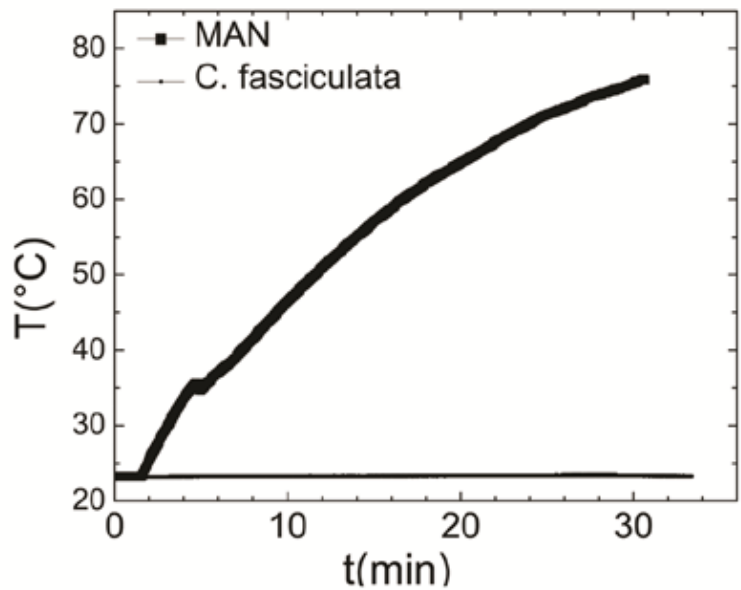

**Figure 6**: Flow citometry results of the '2x2' experiments showed in Figure 2(see text for details). Experiments A); B) and C) showed 89-91% of cell viability, whereas for experiment D) the application of magnetic fields during 30 minutes on magnetically charged cells resulted in only 9% of cell survival.

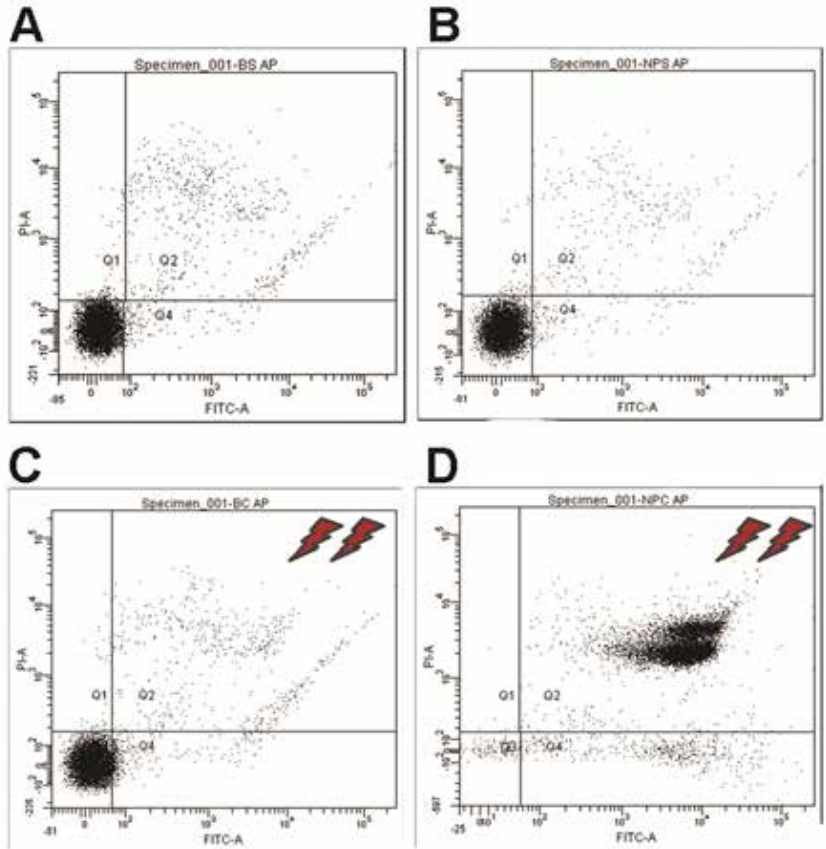

**Supporting information 1:**

Magnetic response at T = 10 K from (a) unloaded cells, (b) the response from co-cultured C. *fasciculata* with MNPs (sample incubated for t = 15 min), (c) the difference between loaded and unloaded cells (b-a), and (d) the pure magnetic colloid. Note that for (d), the pure colloid, the curve was divided by $1.35 \times 10^4$ to fit the same scale than the magnetic signal from loaded cells (fig. S1.c). To calculate the amount of magnetic material $m_{mag}$ incorporated by the cells, the $M_S$ values from the pure colloids and from the MNP-loaded cells were used as $m_{mag} \left[ \frac{g}{cell} \right] = \frac{M}{M_S \times Number\ of\ cells}$. The number of MNPs per single cell was estimated from the known average particle diameter.

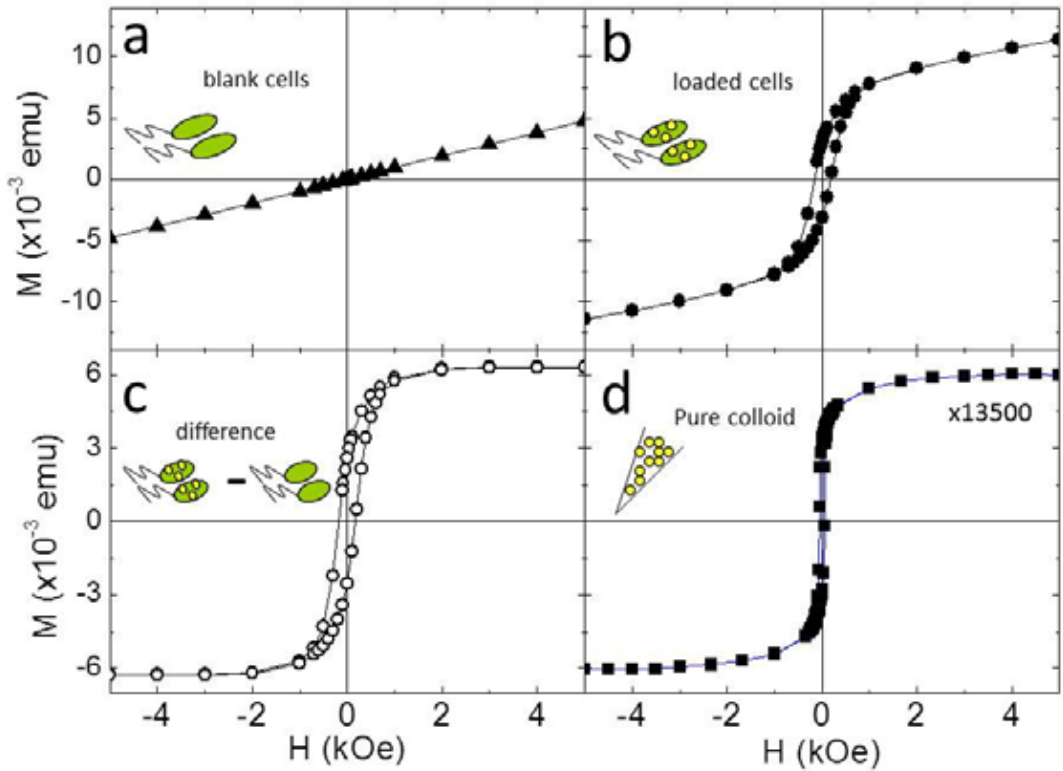

**Supporting information 2:**

Movies corresponding to the 2 x 2 design for evaluating the effect of MNPs and TVMF on *C. fasciculata.*

a) File "a.mov": cells without MNPs not submitted to magnetic fields;

(http://www.youtube.com/watch?v=rZmw8NhsDt0&feature=youtu.be)

b) File "b.mov": cells with MNPs without magnetic field application;

(http://www.youtube.com/watch?v=1wcwnadAjTY&feature=youtu.be)

c) File "c.mov": application of magnetic fields on unloaded cells

(http://www.youtube.com/watch?v=LfIlpnPJ5bk&feature=youtu.be)

and;

d) File "d.mov": application of magnetic fields on MNP-loaded cells.

(http://www.youtube.com/watch?v=EiTx28AZ1gg&feature=youtu.be)